\documentclass[aps,prl,twocolumn,groupedaddress,showpacs]{revtex4}
\usepackage{graphicx}
\begin{document}

\title{Magnetization dynamics in the single-molecule magnet Fe$_8$
under pulsed microwave irradiation}

\author{K. Petukhov,$^{1,2}$ S. Bahr,$^1$  W. Wernsdorfer,$^1$ A.-L. Barra$^2$ and V.Mosser$^3$}

\affiliation{
$^1$Institut N\'eel, associ\'e \`a l'UJF, CNRS, BP 166, 38042 Grenoble Cedex 9, France\\
$^2$Grenoble High Magnetic Field Laboratory, CNRS, BP 166, 38042 Grenoble Cedex 9, France\\
$^3$Itron France, 76 avenue Pierre Brossolette, 92240 Malakoff , France}

\begin{abstract}
We present measurements on the single molecule magnet Fe$_8$ in
the presence of pulsed microwave radiation at 118~GHz. The spin
dynamics is studied via time resolved magnetization experiments
using a Hall probe magnetometer. We investigate the relaxation
behavior of magnetization after the microwave pulse. The analysis
of the experimental data is performed in terms of different
contributions to the magnetization after-pulse relaxation. We
find that the phonon bottleneck with a characteristic relaxation
time of $\sim 10-100$~ms strongly affects the magnetization
dynamics. In addition, the spatial effect of spin diffusion is
evidenced by using samples of different sizes and different ways
of the sample's irradiation with microwaves.

\end{abstract}


\maketitle

\section{Introduction}

Single molecule magnets (SMMs) have attracted much interest in
recent years because of their unique magnetic properties. Having a
regular structure, a well defined spin ground state and magnetic
anisotropy they exhibit quantum phenomena even at macroscopic
scales.\cite{novak:1995,friedman:1996,thomas:1996} Features such as quantum
tunneling between spin states, interference between tunneling
paths or blocking of the spin orientation at very low temperature
show the quantum nature of SMMs.
\cite{barco:2004,sorace:2003,wernsdorfer:science1999,wernsdorfer:2002a,wernsdorfer:2000}
In addition SMMs are supposed to be good candidates for data
storage or quantum computing. \cite{leuenberger:2001}

Recent works in the field of SMMs focused on spin dynamics and
interactions with millimeter-wave radiation. The aima are to control
the spin orientation in the sample and to selectively induce 
transitions between spin states. The crucial point for any
application of SMMs is the knowledge of the spin relaxation time
and the spin decoherence time. Therefore, various experiments have
been performed in studying spin dynamics in SMMs in the presence of
microwaves. Most measurements are based on standard electron paramagnetic resonance (EPR) techniques
\cite{zipse:2003} or on optical spectroscopy \cite{mukhin:2001},
while others are based on magnetization measurements of the sample. In measuring the
absorption of the microwaves via the decrease of magnetization
we can obtain information about both the magnetization of the sample and
EPR-like spectra.
This technique also allows the precise control over the excitation of the sample and makes it
possible to quantify the nonresonant heating.\cite{petukhov:2005}
The magnetization sensor can be either a Hall magnetometer
\cite{sorace:2003,bal:epl2005}, a micrometer sized
superconducting quantum inference device (SQUID) \cite{wernsdorfer:2004}, a
standard SQUID \cite{cage:2005} or an inductive pickup loop
\cite{bal:2004}. Differences in these techniques lie mainly in the
rapidity and sensitivity of the measurement, in the possibility of
applying magnetic fields, and in the compatibility with
microwaves.

In this paper we study the spin dynamics of the single-molecule
magnet $Fe_{8}O_{2}(OH)_{12}(tacn)_{6}$, hereafter called Fe$_8$.
This molecule contains eight Fe(III) ions with spins s=5/2. These
spins are strongly superexchange coupled forming a spin ground
state $S=10$ and the spin dynamics can be described assuming the
giant spin model by an effective Hamiltonian \cite{barra:1996}
\begin{equation}
\mathcal{H} = - DS_z^2 + E(S_x^2 -S_y^2) + \mathcal{O}(4) -
g\mu_B\vec{S}\cdot\vec{H}
\end{equation}

\begin{figure}[b]
\includegraphics[width=3.0in]{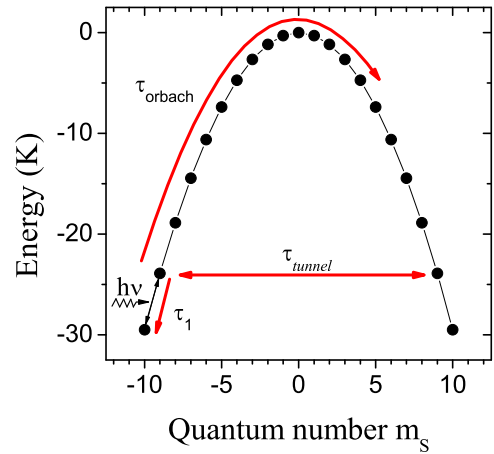}
\caption{\label{barrier} Spin states and energy barrier of the
Fe$_8$ system. For an excited spin there are various processes
relevant for spin dynamics.}
\end{figure}

$\vec{H}$ is the applied magnetic field, $\mathcal{O}(4)$ contains
fourth order terms of spin operators and $g\approx 2$ represents
the gyromagnetic factor. The anisotropy parameters $D=0.275$~K and
$E=0.046$~K have been determined by various experimental methods.
\cite{wernsdorfer:science1999} Classical EPR techniques, frequency
domain magnetic resonance spectroscopy and neutron spectroscopy
are complementary methods and give similar results.
\cite{barra:1996,mukhin:2001,caciuffo:1998,park:2002} The
nondiagonal terms in the Hamiltonian are responsible for the
tunneling processes between spin states, whereas $D$ defines the
anisotropy barrier of approximately 25~K as can be seen in
Fig.~\ref{barrier}.

In terms of the spin dynamics the giant spin model reveals various
relaxation processes that are important for the evolution of the
spin system in the time domain. As sketched in Fig.~\ref{barrier} 
the main parameters of the spin system are the spin
relaxation time $\tau_{1}$ (time scale $\sim 10^{-7}~s$), the
excitation over the barrier by a thermally activated multistep
Orbach process with time constant $\tau_{orbach}$ (time scale
$\sim 10^{-8}~s \times e^{\Delta E/k_{\rm B}T}$) where $\Delta E$
is the barrier height) and the tunnel probability between
degenerated states with time constant $\tau_{tunnel}$ (time scale
$\sim 10^{4}~s$ for the ground state
tunneling).\cite{barra:1996,sangregorio:1997} The use of a large
crystal of the single molecule magnet Fe$_8$ and, in consequence,
the interactions between molecules make it necessary to introduce
spin-phonon and spin-spin interactions. Effects such as spin
decoherence (typical time scale $\tau_ {2}\sim 10^{-9}~s$), phonon
bottleneck (typical time scale $\tau_{ph}\sim 10^{0}~s$) or spin
diffusion ($\tau_{diff}$) have to be taken into
account for a complete description of the spin
dynamics.\cite{Chiorescu:PRL2000,abragam}

In this paper, a series of measurements on the SMM Fe$_8$ is
presented investigating the relaxation of magnetization on 
millisecond and microsecond scales. In Section II we describe the
experimental setup and the various experimental conditions. In
Section III we present the experimental data that will be
discussed in Section IV. Finally in Section V, we give some
concluding remarks.

\section{Experimental techniques}

\subsection{General setup}
The measurements are performed using a commercial 16~T
superconducting solenoid and a cryostat at low temperatures in the
range of 1.4~K to 10~K with temperature stability better than
0.05~K. The magnetization of the Fe$_8$ sample is measured by a
Hall magnetometer. 
The Hall bars were patterned by Thales Research and Technology (Palaiseau),
using photolithography and dry etching, in a delta-doped AlGaAs/InGaAs/GaAs
pseudomorphic heterostructure grown by Picogiga International using
molecular beam epitaxy (MBE). A two-dimensional electron gas is induced in
the 13 nm thick $In_{0.15}Ga_{0.85}As$ well by the inclusion of a Si delta-doping
layer in the graded $Al_xGa_{1-x}As$ barrier. All layers, apart from the quantum
well, are fully depleted of electrons and holes. The two-dimensional electron gas density $n_s$
is about $8.9 \times 10^{11}~cm^{-2}$ in the quantum well, corresponding to a sensitivity
of about 700~$\Omega$/T, essentially constant under $-100~^\circ$C. 
The sample is placed on top of the 10~$\mu$m $\times$ 10~$\mu$m Hall junction
with its easy axis approximately parallel to the magnetic field of
the solenoid. The three samples used in our experiments
(150~$\times$~100~$\times$~30~$\mu$m$^3$,
160~$\times$~180~$\times$~100~$\mu$m$^3$ and
680~$\times$~570~$\times$~170~$\mu$m$^3$) are exposed to microwave
radiation. Microwaves are generated by a continuous wave (cw),
mechanically tunable Gunn oscillator with a nominal output power
of 30~mW and a frequency range of 110~GHz to 119~GHz. Pulses are
generated using a SPST fast PIN diode switch (switching time less of
than 3~ns) triggered by a commercial pulse generator. An oversized
circular waveguide of 10~mm diameter leads the microwaves into the
cryostat. In some of our experiments we use transition parts from
oversized circular-to-rectangular WR6 waveguides. In other
experiments we use only a cylindrical cone as an end piece of the
circular waveguide that is right in front of the irradiated
sample. The different configurations of the coupling of the
microwaves to the sample that are investigated and compared are
explained in the next paragraph. The cryostat is filled with
exchange gas that thermodynamically couples the sample to the bath
and allows a rather fast heat exchange. As the signal of the
Hall magnetometer is in the range of a few microvolts a low-noise
preamplifier is used in order to increase the signal-to-noise
ratio in the rather long coaxial cables. Finally, the signal
acquisition is done by a fast digital oscilloscope having a bandwidth 
of 1~GHz and 10~G/s sample rate and is done by taking an
average over typically 32 frames.

\subsection{Coupling of the microwaves to the sample}

\begin{figure}
\includegraphics[width=3.0in]{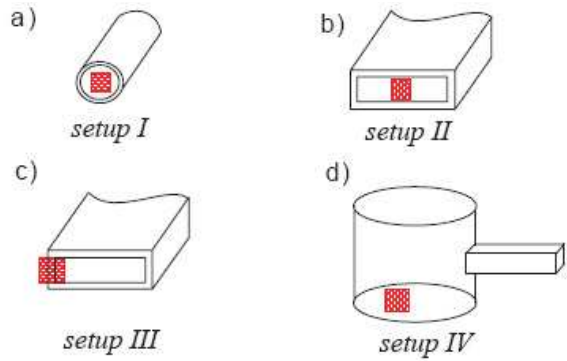}
\caption{\label{wguides} Different configurations of irradiating
the sample with microwaves. We used several waveguides in order to change the electromagnetic environment of the sample,
i.e. the coupling of the microwaves to the sample.}
\end{figure}

\subsubsection{Conical waveguide}
The simplest method of coupling microwaves to the sample is
provided by a conical waveguide focusing microwaves from the
oversized circular waveguide to the sample. In our experiments a
cone with an opening of 4~mm in diameter is used. This setup is
sketched in Fig.~\ref{wguides}a and will be denoted hereafter as 
\emph{setup I}. The conical waveguide has the advantage that it
conserves the polarization of the passing light, thus allowing
experiments depending on the angle of polarization or experiments
with circularly polarized microwaves. Due to the large dimensions of
the waveguide compared to the wavelength the propagation of the
microwaves can be considered as quasioptical and the attenuation
of microwave power is rather small.

\subsubsection{Rectangular waveguide}
Compared to the conical end piece of the waveguide a rectangular
waveguide can focus the microwaves even better. The opening of a
typical D-band WR6 waveguide is 1.7~mm~$\times$~0.8~mm, thus the
cross section is nine times less than that in the case of the conical
waveguide. However, the attenuation using the rectangular waveguide
is fairly large, especially because of the transition part
between oversized cylindrical and rectangular waveguides. At the
end of the rectangular waveguide the field distribution is well
defined. This feature allows us to irradiate the crystal in
different ways. It can be put either in the central region of the
waveguide (Fig.~\ref{wguides}b, \emph{setup II}), or at the edge
of the waveguide, in order to partially irradiate the crystal
(Fig.~\ref{wguides}c, \emph{setup III}). When irradiated partially, 
the magnetization of the crystal is measured with a
Hall sensor at the nonirradiated side of the crystal. This method
allows us to point out the importance of the inhomogeneous distribution
of magnetization in the sample and spin diffusion processes that
take place in large samples.

\subsubsection{Microwave resonator}

In some of our experiments a cylindrical cavity made of copper
with detachable end plates is used, and the sample is placed on top
of one of the end plates (Fig.~\ref{wguides}d, \emph{setup IV}).
The inner diameter of the resonator is 10.10~mm and the height is
5.5~mm. A standard WR6 waveguide is coupled to the sidewall of
the cavity by a coupling hole. The microwaves from the waveguide
enter into the cavity at half height and can excite various modes
obeying the selection rule $TE_{odd,*,*}$ and $TM_{even,*,*}$
(Table~\ref{modes}). All the modes have zero tangential electrical field
on the end plates. The magnetic field has one or more
maxima on the end plates according to the mode and the direction of
the magnetic field is always radial. The sample is mounted on one
of the end plates in such a way that the magnetic field in the cavity
and the easy axes of the sample are parallel.

\begin{table}[htb]
\begin{tabular}{|c|c|}
\hline
Resonator Mode & Frequency [GHz]\\
\hline
$TM_{413}$ & 108.631 \\
$TE_{114}$ & 110.471 \\
$TM_{014}$ & 111.436 \\
$TM_{033}$ & 114.137 \\
$TE_{314}$ & 116.097 \\
$TM_{214}$ & 119.410 \\
$TE_{124}$ & 120.176 \\
$TM_{024}$ & 120.932 \\
\hline
\end{tabular}
\caption[modes]{Possible modes in a perfect cylindrical cavity in
the range of 108~GHz to 121~GHz and for the dimension of the cavity specified in the text. 
Due to perturbations inside the
cavity the calculated frequencies might differ from the
theoretical ones.} \label{modes}
\end{table}

By using a resonator we expect the amplitude of ac magnetic field
to increase by a few orders of magnitude and thus allow better
excitation of the sample even with very short pulses ($<$
10~$\mu$s). The Q-factor for resonant modes is numerically
calculated to be in the range of $10^2$ to $10^3$ depending on the
mode. Therefore, the resonant modes are expected to have a full
width at half maximum in the order of $0.1$ to $1$~GHz.
Consequently we expect each mode to exist in a rather broad
frequency band. The modes should be present in the resonator even
when the microwave frequency does not exactly match the calculated
resonance frequency.

In the experiments two slightly different end plates are used. In
the first case an \emph{unprotected} Hall bar with a sample on top
is directly glued on top of the end plate. The position of the
sample is about 2~mm off center of the end plate. In respect to the
size of the cavity the Hall sensor and the sample represent a
perturbation of the resonator that might be non-negligible. In
consequence the frequencies for the different modes might slightly and inhomogeneously shift 
according to the calculated frequencies. Nevertheless the density of modes between 109~GHz and
120~GHz should remain rather high.

In the second case in order to perturb the cavity as weak as
possible we \emph{protected} the Hall bar with a copper foil. The
position of the sample in this case is about 1~mm off center of
the end plate. A small hole is drilled into the end plate and the
Hall sensor is placed into the hole and is finally coated with
a thin copper foil (thickness of 10~$\mu$m). Thus only the sample
placed on the copper foil is directly exposed to the
electromagnetic field inside the cavity whereas the Hall sensor
and the cables are outside the resonator. In this setup the
Hall sensor is protected from the microwave radiation by the thin
copper foil, however the sensitivity in measuring the sample's
magnetization is expected to be weaker as in the unprotected case.

\section{Measurements}

\subsection{Magnetometry combined with microwaves}

\begin{figure}
\includegraphics[width=3.0in]{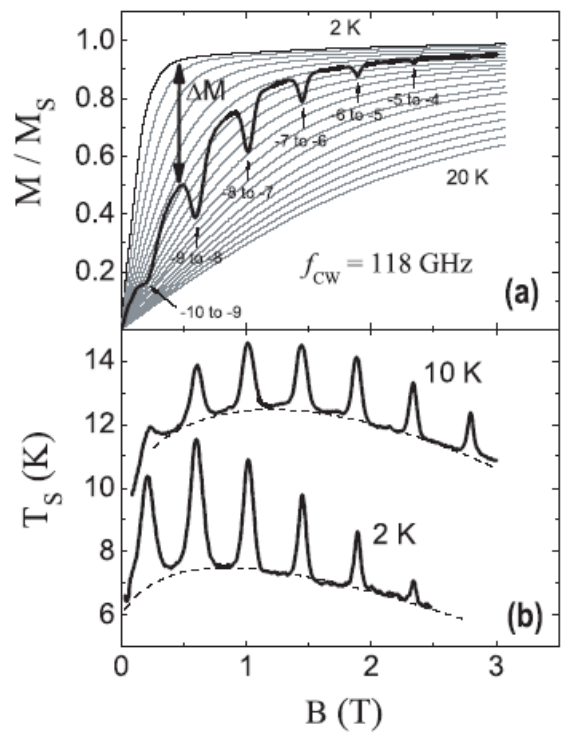}
\caption{\label{CW_stuff}(a) Magnetization of Fe$_8$ as a function
of magnetic field. The curves are normalized to the saturation
magnetization value $M_S$. The thin solid curves represent the
magnetization measured without microwaves at temperatures from 2~K
to 20~K in 1~K steps. The thick solid curve represents
magnetization curve in presence of cw microwaves of frequency
$f_{CW}=118$~GHz measured at 2~K. The thick arrow represents the
magnetization difference $\Delta M$ between data taken with and
without microwaves at 2~K. (b) Spin temperature $T_{\rm S}$
vs applied field $B$, calculated using the mapping
procedure described in the text. The dashed curves depict
off-resonance absorption.}
\end{figure}

When the sample of SMM placed in the magnetic field $B$ is exposed
to the cw microwaves, the magnetization curves
show resonant absorption dips, similar to EPR spectroscopy
spectra. The absorption of microwave radiation takes place at
certain field values at a given frequency, when the microwave
frequency matches the energy difference between two neighboring
energy states, thus the allowed transitions are $\Delta m_s=\pm
1$. The populating of the upper levels (see Fig.~\ref{barrier})
reduces the net sample's magnetization $M$, the change of which
$\Delta M$ can be detected via Hall voltage measurements. If the
applied magnetic field is ramped while the microwaves are applied,
the obtained magnetization spectra clearly show a series of nearly
evenly spaced absorption dips, which can be easily attributed to
the appropriate transitions, as shown in Fig.~\ref{CW_stuff}a by
the thick solid curve. This curve is placed on the top of "pure"
magnetization curves, i.e. the curves measured without microwaves,
depicted by the thin solid curves in Fig.~\ref{CW_stuff}a. These
reference curves were measured at different cryostat temperatures
in the range from 2~K (top curve) to 20~K (bottom curve) with 1~K
incremental step. As can be seen from Fig.~\ref{CW_stuff}a, the
2~K magnetization data measured without microwaves and
the magnetization data measured in the presence of microwaves do not
match: the latter curve lies much below the first curve, as the
temperature during the microwave experiment would be higher
compared to that of the pure magnetization experiment. The difference
between the two curves is denoted by the magnetization difference
$\Delta M$, which is a good measure of the amount of microwave
radiation absorbed by the spin system.

\subsection{Spin temperature}
Although the difference of magnetization $\Delta M$ can be used to
qualify the amount of absorbed microwave photons, it is rather
inconvenient to speak in terms of relative units of $\Delta M$.
Another more significant complication in the use of $\Delta M$ for
quantitative characterizations concerns the loss of sensitivity of
$\Delta M$ close to zero field. As the magnetic field $B$ goes to
zero, the magnetization also goes to zero, and hence the
sensitivity of detection of absorption peaks goes to zero as well.
Therefore, we need to perform a transformation of the
magnetization to a physical quantity which does not depend on the
magnetic field $B$.

Such a quantity called \emph{spin temperature} was explicitly
introduced in our earlier paper~\cite{petukhov:2005} as a perfect
measure of the amount of microwave radiation absorbed by an SMM
spin system. The concept of spin temperature can be easily
understood from Fig.~\ref{CW_stuff}a. We can map the magnetization
curve (magnetization spectrum) obtained under the use of microwaves
onto underlying reference magnetization curves, measured at
different cryostat temperatures without microwaves. For each
magnetization point of the absorption spectra one finds, at the
corresponding field $B$, the temperature $T_{\rm S}$ that gives
the same magnetization measured without microwave
radiation~[Fig.~\ref{CW_stuff}a]. The temperatures in between the
reference magnetization curves are obtained with a linear
interpolation. A typical result of such a mapping is depicted in
Fig.~\ref{CW_stuff}b. $T_{\rm S}$ can be called the spin
temperature because the irradiation time is much longer than the
lifetimes of the energy levels of the spin system which were found
to be around 10$^{-7}$ seconds~\cite{Wernsdorfer:EPL2000}. The
phonon relaxation time $\tau_{ph}$ from the crystal to the heat
bath (cryostat) is much longer, typically between milliseconds and
seconds~\cite{Chiorescu:PRL2000}. The spin and phonon systems of
the crystal are therefore in equilibrium.

Figure~\ref{CW_stuff}b shows spin temperature data calculated for
the magnetization measurements at cryostat temperatures of 2~K and
10~K, performed at frequency of cw microwaves of $f_{CW}=118$~GHz.
From Fig.~\ref{CW_stuff}b we can conclude that the obtained spin
temperatures $T_{\rm S}$ are much larger than the cryostat
temperature $T$. This is associated with a strong heating of the
spin system. This effect is more prominent at lower $T$: the
baseline of 2~K $T_{\rm S}$ spectrum is around 7~K, while 10~K
$T_{\rm S}$ spectrum's background is very close to the nominal
cryostat temperature $T=10$~K, as depicted by the dashed curves in
Fig.~\ref{CW_stuff}b. We also see that both backgrounds are not
flat, but are magnetic field dependent. The presence of the spectra's
nonflat background is due the presence of off-resonance
absorption of microwaves, which takes place between the resonant
absorption peaks. The nonresonant (or background) absorption is
modulated in the following way: it has larger contribution where
the resonant absorption has larger spectral weight, i.e. higher
peaks of $T_{\rm S}$. Interestingly, the off-resonance absorption
was also evidenced in the EPR spectra of SMMs, but its origin 
remains undiscussed~\cite{park:2002,Hill:PRB02}. On the top of the
nonresonant background one can see a perfect EPR-like absorption
spectra, and the $T_{\rm S}$ peak positions exactly match the
magnetic field values, corresponding to $|\Delta m_S|=1$
transitions (see Fig.~\ref{CW_stuff}).

\subsection{Pulsed microwave measurements}

\begin{figure}
\includegraphics[width=3.4in]{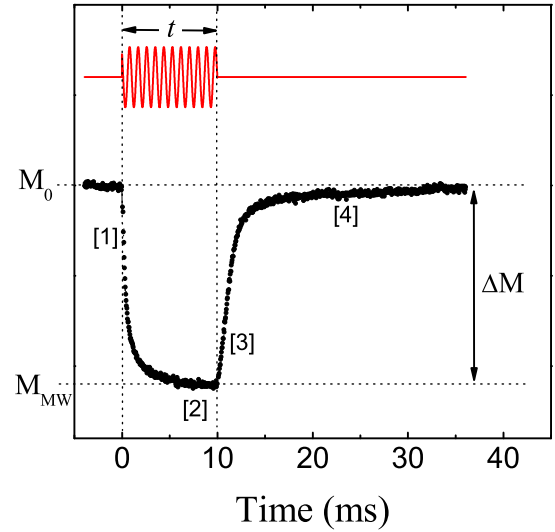}
\caption{\label{OSC}Typical oscillogram of a pulsed experiment.
The magnetization was measured as a function of time for a
microwave pulse length of $t=10$~ms at temperature $T$ = 10 K.}
\end{figure}

Another way to perform $\Delta M$ measurements in order to
calculate the spin temperature $T_{\rm S}$ is to utilize a pulsed
microwave (PW) radiation~\cite{petukhov:2005}. This method gives
direct information about $\Delta M$ at a given magnetic field value
$B$, at a given temperature $T$, and at a given microwave power, a
measure of which is a pulse length $t$. This advanced method also
drastically reduces the heating of the sample with microwaves,
since the repetition rate of microwave pulses (typically 200~ms in
our experiments) is much larger than the pulse length values,
typically $t \leq 10$~ms. Restoration of the after-pulse
magnetization to the equilibrium value $M_0$ normally takes less
than 100~ms, as can be seen in Fig.~\ref{OSC}.

The scheme of possible pulsed experiment is depicted in
Fig.~\ref{OSC}. The top part of Fig.~\ref{OSC} schematically shows
microwave pulse of duration $t=10$~ms, and the bottom part of the
figure shows time-resolved development of magnetization data
collected during such pulsed experiment. The magnetization before
and at the end of the pulse has values $M_{0}$ and $M_{\rm MW}$,
respectively. The difference between the unperturbed magnetization
value $M_{0}$ and magnetization at the final edge of the pulse
$M_{\rm MW}$ (i.e. the height of the magnetization response)
$\Delta M= M_{\rm MW}-M_{0}$ is identical to the magnetization
difference $\Delta M$ defined for the cw microwaves case, as
graphically explained by Fig.~\ref{CW_stuff}a. Thus, having a set
of reference magnetization curves, shown by thin curves in
Fig.~\ref{CW_stuff}a, and magnetization difference $\Delta M$
defined from the PW measurements, as shown in Fig.~\ref{OSC}, one
can perform spin temperature $T_{\rm S}$ calculations. Such
calculations for Fe$_8$ have been performed in our previous work
for the PW configuration and it has been shown that obtained spin
temperatures $T_{\rm S}$ are much closer to the cryostat
temperature $T$ than that for cw experiments~\cite{petukhov:2005}.
The linewidths and shapes of PW $T_{\rm S}$ spectra depend on the
pulse length, but in general the peak positions of cw and PW
configurations are identical. In contrast to the cw experiments,
the PW method can successfully resolve absorption peaks near zero
field~\cite{petukhov:2005}.

Unlike cw measurements, PW magnetization profiles contain 
information not only about $\Delta M$ but also about the magnetization
dynamics. Let us assume that the applied magnetic field $B$ and the
microwave frequency $f_{PW}$ do match the resonance condition,
i.e., the in-resonance microwave pulse is applied. Since the sample's
magnetization is connected to the spin state level occupancy of
SMMs, the magnetization dynamics should be connected to the
level's lifetime. The spin-spin relaxation time $\tau_2$ is
usually much shorter than the spin-phonon relaxation time $\tau_1$; if
$\tau_1$ obtained through magnetization dynamics is short enough,
it can determine the upper limit of $\tau_2$. Finally, since there
is an increase of the sample's temperature due to the heating with
microwaves, the phonon relaxation time $\tau_{ph}$ from the
crystal to the heat bath (cryostat) can be systematically studied
from the magnetization "cooling" after long enough pulses in PW
experiments. Thus, the detailed consideration of the time-resolved
magnetization profile, depicted in Fig.~\ref{OSC}, might provide
information about the $\tau_1$, $\tau_2$, and $\tau_{ph}$
relaxation times.

Let us consider the magnetization behavior during a PW experiment
in detail in Fig.~\ref{OSC}. At the beginning of the pulse, the
magnetization rapidly decreases (region [1]) and starts to
saturate (region [2]) until the end of the pulse. We need to note
that a complete saturation is observed only for rather long pulses
of several seconds. After the microwave pulse is switched off, the
magnetization restores back to the equilibrium value $M_{0}$. At
the beginning of its restoration the magnetization increases
rapidly (region [3]), several millisecond later magnetization
increase changes to the slower behavior (region [4]), until it
levels out at $M_{0}$. This slow restoration lasts long, up to a
hundred of milliseconds, but we were able to follow it completely,
since the typical repetition time of pulses was 200~ms. This
brings us to the conclusion, that region [4] might comprise the
information about the cooling of the sample after the microwave
pulse, i.e. the phonon relaxation time $\tau_{ph}$ from the
crystal to the heat bath (cryostat). Exactly this relaxation time
is typically of the order of magnitude of several tens of
milliseconds up to seconds~\cite{Chiorescu:PRL2000}. The
fast-running beginning of region [3] could contain the
longitudinal relaxation time $\tau_1$ (typically $\sim
10^{-7}$~seconds~\cite{Wernsdorfer:EPL2000}). There was another
interesting observation in region [3] of the magnetization curve
in Fig.~\ref{OSC}: during some of our experiments we have observed
that right after the pulse was switched off the magnetization continued
to decrease for some time and only then started to increase to the
equilibrium value. Similar behavior of magnetization after
microwave pulses in Fe$_8$ was observed in the recent work of Bal
\textit{et al.}~\cite{bal:epl2005}. Below we will explicitly
investigate such an \emph{overshooting} of magnetization after the
microwave pulse. In principle, regions [1] and [2] are very
similar to the regions [3] and [4], correspondingly. The problem
with the use of this part of magnetization evolution curve is that
regions [1] and [2] are limited by the pulse duration, and
therefore it seems to be problematic to estimate relaxation times
from this part of the data, especially the long-lasting
$\tau_{ph}$. In this paper, we pay attention to the
time-resolved behavior of the magnetization after the pulse in a
PW experiment, i.e. to regions [3] and [4] as determined in
Fig.~\ref{OSC}.

\subsection{The model}
\begin{figure}
\includegraphics[width=3.0in]{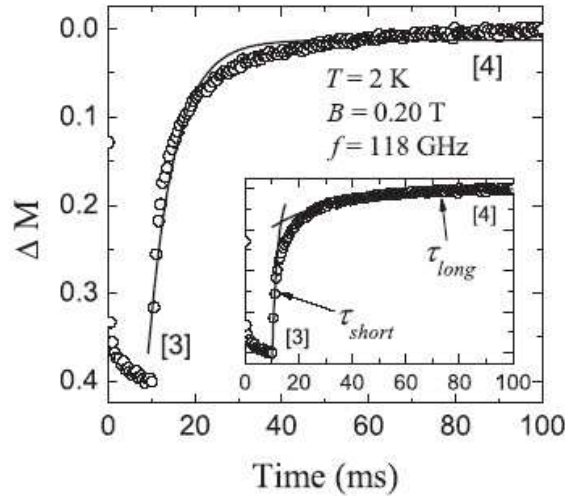}
\caption{\label{fits}Fit of magnetization restoration data (open
circles) with a single exponential relaxation (solid line). The inset
shows the fit with two exponents: fast relaxation with relaxation time
$\tau_{short}$ and cooling with relaxation time $\tau_{long}$.
Note that the depicted magnetization data curve contains only 2\% of
the experimental points taken during the single experiment, i.e.
only every 50th point is shown.}
\end{figure}

First, in order to analyze the behavior of the magnetization after
the microwave pulse, we have tried to fit the magnetization data
in regions [3] and [4] by a single exponential relaxation. We have
found that in many cases a single exponential description was
unsatisfactory, as shown in Fig.~\ref{fits}. This is not
surprising in the framework of consideration concerning different
relaxation times given above. Indeed, the $\tau_1$ relaxation
time, which can be found right after the pulse is much shorter
than the cooling $\tau_{ph}$ relaxation time, which can be a major
contribution in the tail of magnetization restoration. Therefore,
we have separately considered two different regions of the
magnetization restoration curve, as depicted in the inset of
Fig.~\ref{fits}. Firstly, we have assumed that the magnetization
data, obtained right after the pulse was switched off (typically, within
the time frame of 10-20\% of the pulse length), could contain the
information about the spin-lattice $\tau_{1}$ relaxation time.
This data can be described by a fast exponential relaxation, and
we will denote the corresponding relaxation time by
$\tau_{short}$, as shown in the inset of Fig~\ref{fits}. Another
valuable contribution to the overall magnetization restoration
comes from the cooling of the specimen after the microwave pulse,
such a process can be described by the long-lasting relaxation
process with relaxation time $\tau_{long}$, taken later after the
pulse was switched off (typically, 3 to 4 pulse length values later after
the pulse edge until the end of the magnetization data), as
depicted in the inset of Fig~\ref{fits}. If the slow relaxation
$\tau_{long}$ is responsible for the sample's cooling, it can only be
sensitive to the sample size and its thermal coupling to the bath,
with both parameters unchanged during an experimental set.
Therefore, we expect this contribution to be temperature and pulse
length independent. Nevertheless, the slow relaxation
$\tau_{long}$ contribution (i.e. sample thermalization) can become
dominating over the fast relaxation $\tau_{short}$ on increase of
the temperature and/or for very long pulses, since the
$\tau_{short}$ drastically shortens under such conditions and can
be unresolved. In this case, a single exponential relaxation could
be suitable for the magnetization data description and it could
give solely the relaxation time $\tau_{long}$. In our experiments
we avoid such a situation and we carefully adjust the experimental
condition to have two clearly distinguishable regions [3] and [4],
where the uncontroversial analysis by means of $\tau_{short}$  and
$\tau_{long}$ can be performed. This analysis was applied in the
following magnetization relaxation measurements.

\subsection{Relaxation of magnetization}

We have studied the relaxation of magnetization employing
different sample irradiation configurations, described in the
"Experimental setup" part. In all the cases, the applied magnetic
field was set to 0.2~T and the frequency of microwaves during
pulses was 118~GHz. Thus, below we describe the studies of the
magnetization dynamics of the first transition from the ground
state $m_S=-10$ to the first excited state $m_S= -9$, since the
given magnetic field and frequency values match the resonance
condition for Fe$_8$ placed into magnetic field along its easy
axis~\cite{petukhov:2005}. While the temperature and the pulse length were
changed during the PW experiments, shown below, the repetition
time of microwave pulses was always set to 200~ms.

\subsubsection{Measurements with conical waveguide}

\begin{figure}
\includegraphics[width=3.4in]{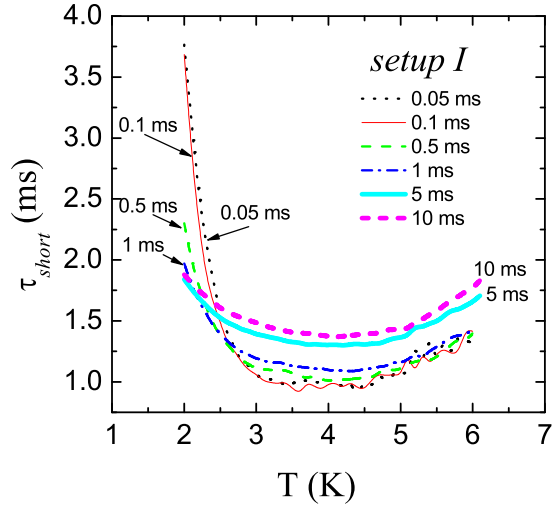}
\caption{\label{Tdeptau}Temperature dependence of relaxation time
$\tau_{short}$ measured for a small sample with a conical
waveguide. The microwave pulses of repetition time of 200~ms and
of different lengths were applied: 0.05~ms (dotted curve), 0.1~ms
(solid curve), 0.5~ms (dashed curve), 1~ms (dashed-dotted curve),
5~ms (bold solid curve), and 10~ms (bold dashed curve). Color
online.}
\end{figure}

Measurements with a conical waveguide are performed on a tiny
sample placed in \emph{setup I} configuration, as shown in
Fig.~\ref{wguides}a. The volume of the sample is 150 $\times$ 100
$\times$ 30 $\mu$m$^{3}$. We have investigated the $\tau_{short}$
and $\tau_{long}$ relaxation times as a function of temperature
$T$ for different values of pulse length. The results obtained
from such PW experiments for the short relaxation time $\tau_{short}$
are depicted in Fig.~\ref{Tdeptau}.

The fast relaxation shows the relaxation time $\tau_{short}$ of the
order of magnitude of 1-1.5~ms. The temperature dependence of
$\tau_{short}$ is not strong, but is clearly pronounced: on warming
from 2~K to approximately 4~K the relaxation time $\tau_{short}$
decreases, reaches its minimum near 4~K, and then smoothly
increases on further warming (see Fig.~\ref{Tdeptau}). The
decrease of relaxation time $\tau_{short}$ on warming from 2~K to
approximately 4~K is much steeper for shorter pulses, while the
$\tau_{short}$ behavior above 4~K is similar for all the pulse
length values.

\begin{figure}
\includegraphics[width=3.0in]{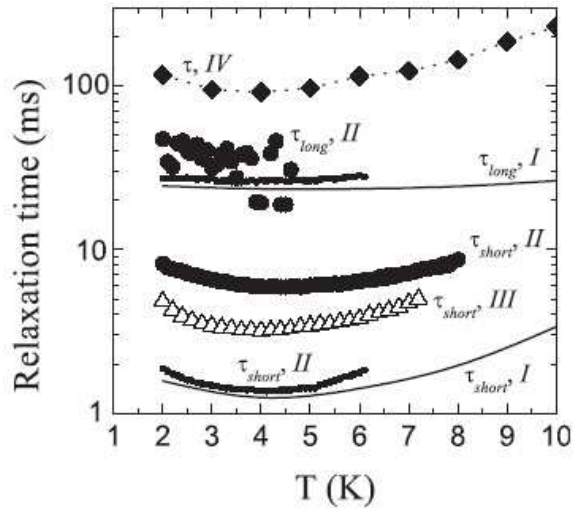}
\caption{\label{relax}Generic plot depicting temperature
dependence of fast relaxation time $\tau_{short}$ and slow
relaxation time $\tau_{long}$ for different sample sizes and
different sample's irradiation configurations: \emph{setup~I}
(solid curves), \emph{setup~II} for a big and a small sample (big
and small solid circles, respectively), \emph{setup~III} (open
triangles), and \emph{setup~IV} (solid rhombus with dotted curve).
Note that the Latin numbers stand for corresponding setup notation.}
\end{figure}

For the same sample and experimental configuration \emph{setup~I},
the slow relaxation time $\tau_{long}$ is approximately one order
of magnitude greater, that is, its value lies around 20~ms. This
long-lasting relaxation time is temperature independent within 5\%
in the measured temperature range, as shown in Fig.~\ref{relax},
where both $\tau_{short}$ and $\tau_{long}$ obtained from the
10~ms data using \emph{setup~I} configuration are shown by solid
curves for comparison. This behavior of $\tau_{long}$ perfectly fits
into the above given explanation of $\tau_{long}$-process as
a cooling of the sample after the microwave pulse. Such a cooling
rate is only defined by the sample thermal coupling to the bath,
and therefore, it is permanent for a given experimental setup.

Figure~\ref{relax} shows the generic plot of relaxation times
$\tau_{short}$ and $\tau_{long}$, obtained from all the
experimental configurations at 10~ms pulse length over 200~ms
pulse repetition time. Such a rather long pulse length was chosen
mainly due to the following reasons. At first, at rather long pulses
the contribution of the $\tau_{long}$-process, i.e. the sample
thermalization after the pulse, becomes valuable, making 
better separation of $\tau_{short}$ and $\tau_{long}$
data intervals possible, and thus both fast and long relaxations can be
estimated and compared for the same system under the same
experimental condition. Secondly, at long pulses, a high
signal-to-noise ratio enables better fitting, even for the limited
regions of experimental data. Finally, as will be shown below,
at such long pulses there are no \emph{overshooting} phenomena
observed for both small and big samples at all the temperatures
applied; with the presence of overshooting, the analysis of
magnetization data in terms of relaxation exponents becomes
controversial.

\subsubsection{Measurements with a rectangular waveguide}

The irradiation of the sample with a piece of the rectangular
waveguide WR6 is advantageous in the sense that the distribution
of electromagnetic field is known at the waveguide cut edge.
Another advantage of the use of the WR6 waveguide is that the area of
its opening (1.36~mm$^2$) is approximately 9 times smaller than
the area of the 4~mm opening of the conical waveguide, so we could
expect a higher density of microwaves exposed to the sample. On
measurements employing rectangular waveguide, \emph{setup~II}
configuration of irradiating the sample with microwave radiation
is used. In this configuration, the sample is placed in the
geometrical center of the waveguide opening, as schematically
shown in Fig.~\ref{wguides}b. \emph{Setup~I} configuration
provides the point of maximal magnetic field for the propagating
TE$_{10}$ mode in a rectangular waveguide~\cite{poole}.

In our magnetization study employing \emph{setup~II}
configuration, we have used two samples of different volumes: a
big sample with a volume 680$\times$570$\times$170~$\mu$m$^3$ and a
smaller sample with a volume of 160$\times$180$\times$100~$\mu$m$^3$.
We will refer these samples hereafter as \emph{big} and
\emph{small}, correspondingly. The measurements were performed
upon irradiation with pulsed microwaves of 118~GHz and at applied
magnetic field of 0.2~T, these conditions correspond to the first
transition from the ground state -10 $\rightarrow$-9 for Fe$_8$
system along the easy axis. We have found that the relaxation time
$\tau_{short}$ for the \emph{small sample} behaves very similar to
the relaxation time $\tau_{short}$ of the sample used with
\emph{setup~I}. Indeed, this relaxation time decreases during the
temperature increase from 2~K to 4~K, where it reaches its minimal
value of around 1.5~ms, as shown by the small solid circles in
Fig.~\ref{relax}. On the consequent increase of the temperature above
4~K, we see the increase of $\tau_{short}$ with temperature $T$.
The $T$-dependence of the fast relaxation parameter $\tau_{short}$
for the \emph{big} sample is qualitatively similar to that of the
\emph{small} sample, the corresponding data is depicted with big
solid circles in Fig.~\ref{relax}. Here, the relaxation time also
reveals a minimum at around 4~K, but the absolute values of
$\tau_{short}$ for the \emph{big} sample are approximately four times
higher, and both dependences can be perfectly scaled one onto
another. The $\tau_{long}$ values for both the \emph{small} and
\emph{big} samples are shown in Fig.~\ref{relax} by the small and big
solid circles, respectively; both these relaxation times lie
around 25-30~ms and they are temperature independent, similar for
the earlier described configuration \emph{setup~I}. This is not
surprising, since in both configurations of coupling of the
samples to the electromagnetic (EM) radiation, the surrounding of
the sample was the same: we have used the same sample holder in
both cases, and the same amount of exchange gas was contained in
the sample volume chamber for the sample's thermalization. This was
not the case when the sample was placed onto the interior wall of
the massive copper cavity resonator.

\begin{figure}
\includegraphics[width=3.0in]{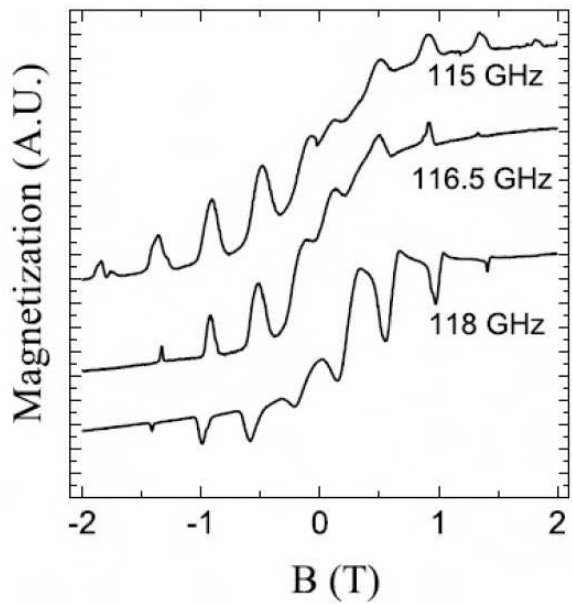}
\caption{\label{cavity}Magnetization loops measured at several cw
microwave frequencies using a cavity. Note, that the curves are
equally scaled.}
\end{figure}

\subsubsection{Measurements with a cavity}

\begin{figure}
\includegraphics[width=3.4in]{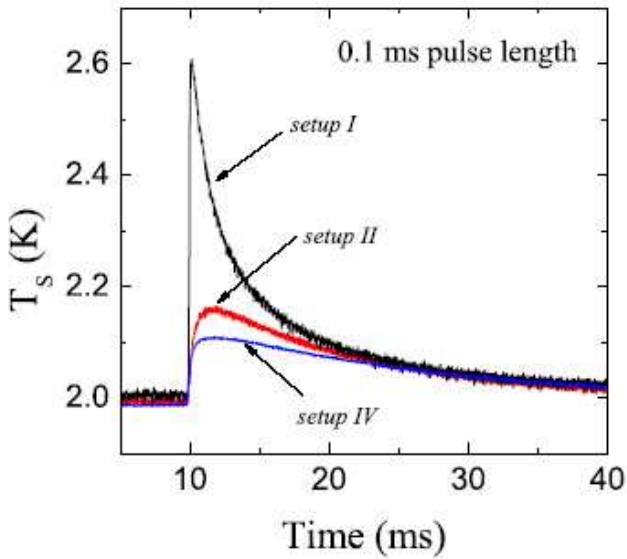}
\caption{\label{ST}Spin temperature calculated during a pulse of
length of 0.1~ms at nominal cryostat temperature $T=2$~K for the
different ways of irradiating sample with microwaves. Color
online.}
\end{figure}

In order to increase the amplitude of the EM radiation exposed to the
sample, we have used the cylindrical copper cavity resonator,
construction details of which are given above. The corresponding
configuration is denoted as \emph{setup~IV} and it is shown in
Fig.~\ref{wguides}c. The use of a microwave cavity assumes the use
of different modes compatible with the cavity geometry and the microwave
frequency. The general problem of the cavity usage is that the
modes of the cavity are coupled to the sample in a different way,
i.e. the electromagnetic environment of the sample inside the
cavity is strongly dependent on the mode and the frequency. This
leads not only to different amplitudes of exposed (and therefore,
absorbed) microwaves, but also to the irregularities of the
absorption spectra. Figure~\ref{cavity} shows the microwave absorption
spectra of magnetization obtained at several frequencies,
separated just 1.5~GHz from each other; the spectra are obtained
by the use of cw microwaves with \emph{setup~IV}. The sample used for
the studies with the cavity has dimensions of
680$\times$570$\times$170~$\mu$m$^3$. Unfortunately, when a
smaller sample is used in the cavity-employed configuration
\emph{setup IV}, the sensitivity is reduced drastically. This is due
to the fact that the sample's change of magnetization is sensed by
the Hall bar separated by a copper foil, as described above.

It can be seen that the modes differ not only in the amplitude of
absorption peaks, but some of them are also highly distorted
(peaks instead of dips, amplitude-phase mixing, asymmetry for
opposite field directions). For the magnetization relaxation
measurements we choose the frequency of 118~GHz, since the mode
working at this frequency shows the largest amplitudes of
absorption peaks in the positive magnetic-field direction; so, all
the measurements presented below are taken at the frequency of
118~GHz.

In general, we find that there is no increase of the EM-field
amplitude exposed on the sample in comparison with the case, when
the sample is irradiated with microwaves without a cavity. The
best way to quantitatively characterize the amount of EM radiation
(photons) absorbed by the sample is to consider the
spin-temperature $T_{\rm S}$ growth due to the exposure of a
microwave pulse. Using the mapping procedure described above, we
convert the magnetization data obtained at a pulse length of 0.1~ms into
$T_{\rm S}$ for the different configurations of coupling of the
sample to the microwaves. The plot, representing spin temperature
$T_{\rm S}$ calculated when no cavity is used (\emph{setup~I} and 
\emph{setup~II}) and in cavity-employed configuration
(\emph{setup~IV}), is depicted in Fig.~\ref{ST}. The plot
presented in Fig.~\ref{ST} shows no evidence of enhancement of
absorption of microwaves, when the resonant cavity is used,
although the best performing mode is chosen for this spin
temperature comparison. Instead, both waveguide-employed
configurations clearly show a better performance. The same sample is
used for \emph{setup~II} and \emph{setup~IV} configurations, while
the smaller sample is used in configuration \emph{setup~I};
details of the sample's size are given above.

The use of a cavity also significantly extends the relaxation of
magnetization after the microwave pulse. This slowing of
magnetization restoration is also clearly visible from the
spin-temperature dynamics, shown in Fig.~\ref{ST}, as compared to
the experiment without a cavity \emph{setup~II} on the same
sample. We have summarized the relaxation parameters when the
cavity was used on the generic plot shown in Fig.~\ref{relax} and
compared these relaxation time values to the parameters, obtained
from the magnetization restoration during pulsed microwave
measurements on the same sample without cavity. The relaxation
time data obtained during cavity-employed experiment are denoted
as "$\tau$, \emph{IV}" on Fig.~\ref{relax}. The obtained relaxation
time $\tau$ is around 100-200~ms, which is one order of magnitude
larger than the slow relaxation time $\tau_2$, obtained in
no-cavity experiments. We cannot unambiguously attribute the obtained
relaxation time $\tau$ to the fast relaxation $\tau_{short}$,
although the data for its calculation were taken right after the
pulse, as what was done for the $\tau_{short}$ definition in no-cavity
setups. We think that the calculated relaxation time $\tau$ 
rather corresponds to the mixture of $\tau_{short}$ and cooling of
the sample thermally coupled to the massive copper cavity, i.e.
the relaxation time $\tau_{long}$.

\subsubsection{Background absorption}

\begin{figure}
\includegraphics[width=3.4in]{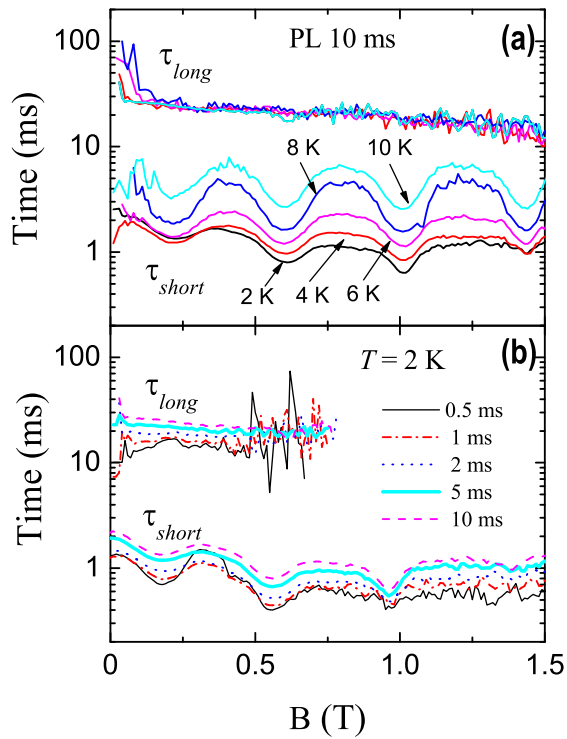}
\caption{\label{backgrnd}(a) Relaxation time $\tau_{short}$ and
$\tau_{long}$ as a function of applied magnetic field in PW
measurements of pulse length of 10~ms and frequency $f=118$~GHz.
The measurements were performed using \emph{setup~I} at several
temperatures. (b) Relaxation time $\tau_{short}$ and
$\tau_{long}$ as a function of applied magnetic field at
temperature $T=2$~K and frequency $f=118$~GHz. The measurements
were performed using \emph{setup~I} at several pulse length
values: 0.5~ms (solid curve), 1~ms (dotted-dashed curve), 2~ms
(dotted curve), 5~ms (bold solid curve), and 10~ms (dashed curve).
The repetition time of the pulses was 200~ms. Color online.}
\end{figure}

All the measurements, described above are performed at a resonance
condition corresponding to the transition from the ground state,
i.e. -10 $\rightarrow$ -9. At 118~GHz, a frequency which is used
for current studies, the appropriate applied magnetic field is
always set to 0.2~T, and thus the resonant condition is fulfilled
for the Fe$_8$ system along the easy axis. Thus, only the resonant
absorption is detected. Nevertheless, as we have mentioned above,
there is also a significant off-resonant, or background,
absorption.

In Fig.~\ref{backgrnd} we present the relaxation time
$\tau_{short}$ and $\tau_{long}$ as a function of the applied magnetic
field. The magnetic field is set in discrete steps from zero field
to 1.5~T with an increment of 0.05~T. These relaxation times are
calculated from the PW measurements performed using configuration
\emph{setup~I} at several temperatures [Fig.~\ref{backgrnd}a] and
at several pulse length values [Fig.~\ref{backgrnd}b].

Both figures show that the long relaxation time $\tau_{long}$
remains pulse length and field independent within the noise
bandwidth and equals to approximately 25~ms; there is no resonance
structure evidenced in $\tau_{long}$ field-dependence. This is
consistent with our consideration of the slow relaxation as a
cooling of the system. Note that above approximately 0.5~T the
magnetization deviation amplitudes are reduced for short pulse
values and the corresponding $\tau_{long}$ curves become very
noisy in Fig.~\ref{backgrnd}b.

$\tau_{short}$ follows the resonance behavior and clearly
pronounced resonant dips can be seen in both figures. We see that
off-resonance and in-resonance relaxation time $\tau_{short}$
values lie within the same order of magnitude (around 1-3~ms),
while $\tau_{long}$ is one order of magnitude larger. For the
pulse length value of 0.5~ms [Fig.~\ref{backgrnd}b], the
difference between the in-resonance value at $H=0.2$~T and
the off-resonance value of $H=0.38$~T is a factor of 2, while for the
pulses with a duration of 10~ms this factor is reduced to 1.2.

\subsection{Magnetization overshooting}
\begin{figure}
\includegraphics[width=3.4in]{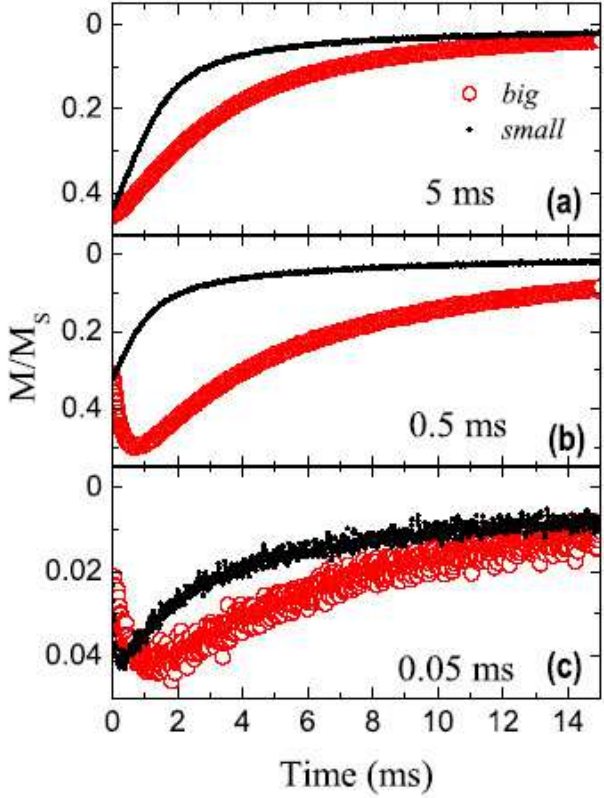}
\caption{\label{over3pulse}Typical plots of PW measurements of
magnetization as a function of time performed on \emph{big} (solid
circles) and on \emph{small} (open circles) samples. The data
plotted are taken at 5~ms (a), 0.5~ms (b), and 50~$\mu$s (c) right
after the microwave pulse was switched off. Color online.}
\end{figure}

\begin{figure}
\includegraphics[width=3.4in]{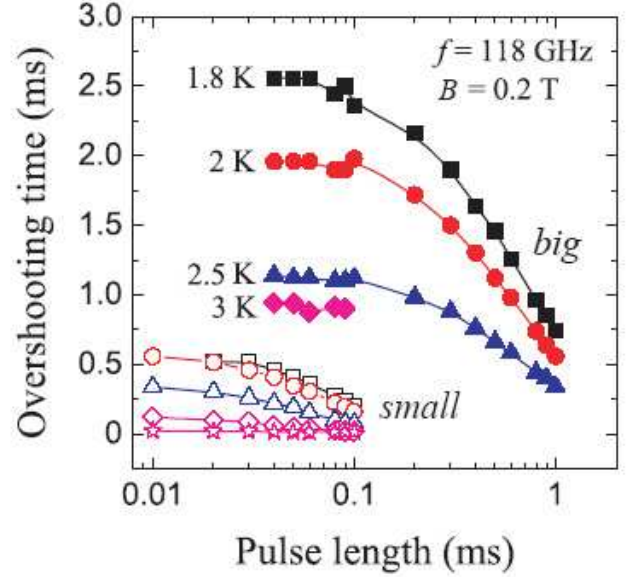}
\caption{\label{over4freq}Position of the magnetization
overshooting minima as a function of pulse length for the big
(solid symbols) and for the small (open symbols) samples. The
measurements were done at temperatures of 1.8~K (squares), 2~K
(circles), 2.5~K (triangles), 3~K (diamonds), and 4~K (stars).
Color online.}
\end{figure}

As we have mentioned above, in some of our PW experiments we
observed an \emph{overshooting} of the magnetization after the
microwave pulse, when the magnetization continued to decrease even
after the pulse is switched off. This phenomenon, for example, can be
clearly seen in the magnetization data recalculated into spin
temperature for \emph{setup~II} and \emph{setup~IV}, as shown in
Fig.~\ref{ST}. A similar effect is also evidenced in the work of Bal
\textit{et al.} \cite{bal:epl2005}. Such an \emph{overshooting},
however, is not observed in magnetization measurements employing
\emph{setup~I}, where we have performed pulsed microwave experiments with pulses of
the length of 10~ms down to 1~$\mu$s. Note that two different
samples are used with \emph{setup~II}, and the volume of the small
sample used in this study is approximately 6 times larger
than the volume of the crystal, used in the measurement with
\emph{setup~I}.

In measurements employing rectangular waveguide, two
configurations \emph{setup~II} and \emph{setup~III} are possible.
In the former configuration, the sample is placed in the geometrical
center of the waveguide opening; in the latter configuration, the
sample is placed at the midpoint of the shortest wall of the
waveguide, as schematically shown in Fig.~\ref{wguides}. Both
configurations provide the points of maximal magnetic field for
the propagating TE$_{10}$ mode in a rectangular
waveguide~\cite{poole}. In \emph{setup~II} configurations, a
sample of any size can be used, while only a sample of large
enough volume can be used in \emph{setup~III}, because a tiny
sample cannot be properly placed at the edge of the waveguide for
partial irradiation with microwaves.

In order to find the nature of the overshooting of magnetization
restoration after the microwave pulse, we employ 
\emph{setup~III} configuration, which allows partial irradiation
of the sample with microwaves. We construct a sample holder, where
\emph{setup~II} and \emph{setup~III} can be used
simultaneously, i.e. two samples can be exposed to the microwaves
at the same time. The change of each sample's magnetization can be
sensed by an individual Hall bar mounted underneath the sample.

For measurements we use two different size samples of Fe$_8$:
one sample, hereafter referred to as \emph{small}, had dimensions of
160$\times$180$\times$100~$\mu$m; and another sample, hereafter
referred to as \emph{big}, had dimensions of
680$\times$570$\times$170~$\mu$m. The \emph{small} sample was
placed in \emph{setup~II} and was entirely irradiated with microwave
radiation. The \emph{big} sample was placed in
configuration \emph{setup~III} and only a part of it was exposed
to the microwaves; the Hall bar was placed under the "dark"
part of the big sample. Both samples were mounted with their easy
axes parallel to the direction of the applied magnetic field set to
the value of 0.2~T, which corresponds to the first transition -10
$\rightarrow$ -9 at the frequency of 118~GHz.

The typical oscillograms of pulsed microwave measurements
performed at temperature $T=2$~K and pulse durations of 5~ms,
0.5~ms, and 50~$\mu$s for the \emph{big} and the \emph{small}
samples are presented in Fig.~\ref{over3pulse}; and the data
plotted are taken right after the microwave pulse was switched off. As seen
in Fig.~\ref{over3pulse}a, after rather long pulses of duration
of 5~ms, no magnetization overshooting is observed for both
the \emph{big} and \emph{small} samples. At ten times shorter pulse
length of 0.5~ms, the magnetization restoration of the
\emph{small} sample reveals no overshooting feature, while
the magnetization of the \emph{big} sample continues to decrease,
reaches the minimum at approximately 0.6~ms after the microwave
pulse is switched off, and only then increases and saturates to the
equilibrium value [see Fig.~\ref{over3pulse}b]. When the
microwaves are applied within the pulses of length of 50~$\mu$s,
the magnetization data of the \emph{big} sample show even more
overshooting: the minimum of the magnetization curve is observed
approximately 1.2~ms after the pulse edge, see
Fig.~\ref{over3pulse}c. At the same time, the \emph{small} sample
magnetization data also show an appearance of overshooting having
its minimum at 0.2~ms after the pulse edge, as can be evidenced
from Fig.~\ref{over3pulse}c.

By performing a series of similar PW experiments in a broader
range of microwave pulse length values and at several helium
temperatures, we obtain a generic plot, depicted in
Fig~\ref{over4freq}. Here, the position of magnetization minima,
i.e. the overshooting time, is plotted as a function of the
applied microwave pulse length values at the resonance condition
of the transition -10$\rightarrow$-9 (118~GHz, 0.2~T). The
measurements are done consequently on the \emph{big} and on
the \emph{small} sample at same temperature and pulse length values.
From Fig.~\ref{over4freq} it can be clearly seen, that the
\emph{big} sample shows well pronounced overshooting already at
pulse lengths of 1~ms and above, while shorter pulses of the
length of approximately 100~$\mu$s are needed in order to observe
measurable overshooting of the magnetization of the \emph{small}
sample. Another interesting finding, which can be concluded from
the dependencies shown in Fig~\ref{over4freq} is that the
overshooting time strongly decreases with the temperature increase.
This temperature dependence is very intense: by increasing the
temperature from 1.8~K to 2.5~K, the overshooting time is reduced
twice in its value. At high enough temperatures the overshooting
feature completely disappears for both samples. The pulse length
dependence of the overshooting time can also be easily understood
in terms of its strong temperature dependence, since the PW
configuration of experiments, as well as that of cw, leads to the
significant heating of the system, as was shown
previously~\cite{petukhov:2005}. Heating with microwaves perfectly
explains why less overshooting is observed at longer pulses than
at shorter pulses.

We have also estimated the relaxation time $\tau_{short}$ for the
\emph{big} sample used in \emph{setup III} configuration. The
corresponding data are plotted by the big open triangles in
Fig.~\ref{relax}. We observe, that the relaxation time
$\tau_{short}$ is very similar to $\tau_{short}$ values obtained
in other cavity-free configurations. In particular, the profile of
the relaxation time $\tau_{short}$ as a function of temperature for
the \emph{big} sample measured with \emph{setup III} is similar to
that of the same sample measured with \emph{setup II} (big solid
circles in Fig.~\ref{relax}). The corresponding absolute values
are very close and the difference between the two curves of 3~ms can
be explained by the partial irradiation of the sample with
microwaves in \emph{setup III}. Thus, only a fraction of molecules
contributes to the magnetization change, and we can consider that
the "effective" size of the sample is smaller.

\section{Discussion}
The magnetization dynamics measurements presented in this work
intend to define some characteristic relaxation times, which
should be taken into account when the spin dynamics of Fe$_{8}$
SMM is considered. In particular, we have investigated
magnetization recovery right after the microwave pulse, where the
spin-phonon relaxation time $\tau_1$ can contribute to the
magnetization relaxation. We have found that the after-pulse fast
relaxation $\tau_{short}$ is typically on the order of magnitude
of several milliseconds, as can be seen in Fig.~\ref{relax}. It
was found that the lower limit for $\tau_{short}$ is $1.4 \cdot
10^{-3}$~s, which is orders of magnitude larger than the
longitudinal relaxation time $\tau_1$, which is expected to be
$\sim 10^{-7}$~s~\cite{Wernsdorfer:EPL2000}. Such an obvious
discrepancy shows that $\tau_1$ contribution to the
$\tau_{short}$-process is not major at the conditions of the
performed experiments. Indeed, the temperature behavior of
$\tau_{short}$ is also incompatible with the expected temperature
behavior of $\tau_{1}$, which should decrease with temperature
growth.

One of the dominant contributions to the $\tau_{short}$ relaxation
can be the phonon-bottleneck effect, which can screen out the 
shorter relaxations, such as $\tau_1$. Within the model described
above, we have performed the magnetization data treatment by means
of the long relaxation time $\tau_{long}$, which is believed to be
characteristic for the cooling of the specimen after the microwave
pulse. The values of $\tau_{long}$ were found to be an order of
magnitude higher than $\tau_{short}$, typically around 30-50~ms.
We have also not evidenced any temperature (see Fig.~\ref{relax})
or magnetic-field (see Fig.~\ref{backgrnd}a) dependence of
$\tau_{long}$. It can be noticed from Fig.~\ref{relax} that
$\tau_{long}$ has a pronounced sample size dependence: for the larger
sample $\tau_{long}$ shown by the big solid circles lies above the
$\tau_{long}$ data for the smaller sample, depicted by the small solid
circles. Another interesting observation is that $\tau_{long}$ has
a prominent power dependence: as can be concluded from
Fig.~\ref{backgrnd}b, the longer pulses provide large
$\tau_{long}$ than shorter pulses. There is nearly a factor of 2
difference between the $\tau_{long}$ data obtained after a pulse
with durations of 0.5~ms and 10~ms. Thus, we attribute the obtained
$\tau_{long}$ relaxation time to the phonon relaxation time from
the crystal to the heat bath $\tau_{ph}$. Our values are in good
agreement with previously published literature
values~\cite{Chiorescu:PRL2000}.

Another observation, which can support the idea that $\tau_{ph}$
is admixed to the $\tau_{1}$ data is that values of $\tau_{1}$
obtained at nonresonant and resonant conditions are rather
similar, as shown in Fig.~\ref{backgrnd}. Although the modulation
due to the resonant absorption can be clearly seen from the data,
the $\tau_1$ values obtained in resonance and out of resonance
differ only by a factor of two. Also, from the plot in
Fig.~\ref{backgrnd}b it can be seen that $\tau_{short}$ and
$\tau_{long}$ relaxation times experience fairly similar power
dependence: an extension of both relaxation times is observed for
longer pulses. This pulse length (or power) dependence of
$\tau_{short}$ can be a plain evidence that a process of cooling
of the crystal contributes to $\tau_{short}$ too.

We have also found another pertaining to time process, which
builds the overall profile of the magnetization restoration curve,
as sketched in Fig.~\ref{OSC}. As it was shown in comparative
experiment on small and big samples (\emph{setup II} and
\emph{setup III}), under certain physical conditions an
overshooting can be observed in magnetization dynamics. The
generic plot depicted in Fig.~\ref{over4freq} shows the mapping of the
occurrence of the overshooting. It is shown in
Fig.~\ref{over4freq} that the sample's size and the sample's
temperature are two factors responsible for the phenomenon of
overshooting in the following way: the larger the sample and the
lower the temperature, the more prominent the overshooting. It
perfectly explains why no overshooting is evidenced in our
previous work~\cite{petukhov:2005} and in investigations employing
\emph{setup I} in this work: we have used a very small sample,
the volume of which was approximately 6 times less than the volume of the
\emph{small} sample used to make the plot in Fig.~\ref{over4freq}.

Such a spatial effect, which also depends on the sample's spin
temperature, can be described by the sample's thermal spin
equilibration. In terms of spin language, this process is known as
\textit{spin diffusion}: for the system of identical spins, where
the level population at one part in the sample is different from
those at other points, the spin flip process will act to make the
population difference uniform throughout the
specimen.\cite{abragam} Thus, spin diffusion creates a uniform
spin temperature throughout the sample. The presence of spin
diffusion is a sequence of the fact that we measure an array of
magnetic molecules, and the spin interactions between them are
presented. Here one can see that the term "single" in the SMM
notation is, to a certain extent, an idealization. In the strict
sense, the spin diffusion is completely inevitable until one
single molecule is measured.

As can be seen from Fig.~\ref{over4freq}, the overshooting time
is comparable to the pulse length. For rather big samples it
can be in the order of magnitude of several milliseconds, which is
already comparable to the values of $\tau_{short}$. Therefore,
experimental conditions should be chosen carefully for such pulsed
microwave experiments. Ideally, one should employ a smallest
possible sample; then even shorter microwave pulses can be
utilized than those depicted in Fig.~\ref{over4freq}.
Nevertheless, to perform microsecond and submicrosecond pulsed
microwave measurements one needs a higher power of microwaves.

As can be seen from Fig.~\ref{ST}, the use of a microwave resonator
cannot serve to reach this goal (\emph{setup IV} in
Fig.~\ref{ST}). The problem is that microwaves can only be guided
to the cavity by a standard rectangular waveguide, for which the
electromagnetic-field distribution of propagating mode is known
and the effective magnetic coupling via coupling hole is possible at
the position of the magnetic-field antinode. Employing such a
configuration, we produce unavoidable losses due to the transition
from the oversized circular waveguide to the WR6 rectangular
waveguide and thus reduce the overall performance of the use of
a cavity. For the same reason the configuration \emph{setup II} is
less advantageous as configuration \emph{setup I}, as shown in
Fig.~\ref{ST}: the use of a circular-to-rectangular transition
leads to high losses. Therefore, there is no gain in the use of better
focusing lower-cross-section rectangular waveguide.

\section{Conclusions}
We have presented the magnetization dynamics experiments employing
magnetization measurements combined with pulsed microwave
absorption measurements. The analysis of the magnetization
dynamics is performed in terms of characteristic exponents, which
describe the fast and slow components of magnetization relaxation.
These exponents are physically connected to different
contributions to the overall magnetization dynamics. We have found
that the spin-phonon relaxation time $\tau_1$ is screened out by
other longer-lasting relaxations. The phonon-bottleneck effect
is probably the major contribution to the magnetization
relaxation, giving a slow relaxation. We have found that the
phonon relaxation time $\tau_{ph}$ is around 30~ms in our
experiments, which is comparable to other studies.\cite{Chiorescu:PRL2000}
We have also evidenced the effect of spin diffusion inside the specimen, which
should be taken into consideration, when after-pulse
magnetization dynamics is analyzed.

Finally, we can propose that more advanced microwave experiments
are needed to resolve the spin-phonon relaxation time $\tau_1$, such
as the "pump and probe" technique employing two frequencies of pulsed
microwaves. But in all cases special care should be taken concerning the
sample's coupling to the microwaves and to the
phonon bath.

\section{Acknowledgments}

We thank R.~Sessoli and L.~Sorace for helpful discussions. The
samples for the investigations were kindly provided by A.~Cornia. This
paper is partially financed by EC-RTN-QUEMOLNA Contract No.
MRTN-CT-2003-504880.

\end{document}